\documentclass[prl,aps,twocolumn,showpacs,floatfix]{revtex4-1}
\usepackage{amsfonts}
\usepackage{amssymb}
\usepackage{hyperref}
\usepackage{graphicx}
\usepackage{dcolumn}
\usepackage{bm,amsmath,verbatim}
\usepackage{mathrsfs}
\usepackage{color}
\usepackage{times}

\usepackage[T1]{fontenc}
\usepackage[latin9]{inputenc}
\usepackage{booktabs}

\providecommand{\tabularnewline}{\\}
\renewcommand\toprule{\hline\hline}
\renewcommand\bottomrule{\hline\hline}

\hypersetup{colorlinks,
linkcolor=blue,          citecolor=blue,        filecolor=blue,      urlcolor=blue           }

\def\be{\begin{equation}}
\def\ee{\end{equation}}
\def\bea{\begin{eqnarray}}
\def\eea{\end{eqnarray}}
\def\bse{\begin{subequations}}
\def\ese{\end{subequations}}

\def\be{\begin{eqnarray}}
\def\ee{\end{eqnarray}}

\begin{document}

\title{Topological Amorphous Metals}
\author{Yan-Bin Yang$^{1}$}
\author{Tao Qin$^{2}$}
\author{Dong-Ling Deng$^{1}$}
\author{L.-M. Duan$^{1}$}
\author{Yong Xu$^{1}$}
\email{yongxuphy@tsinghua.edu.cn}
\affiliation{$^{1}$Center for Quantum Information, IIIS, Tsinghua University, Beijing 100084, People¡¯s Republic of China}
\affiliation{$^{2}$Department of Physics, School of Physics and Materials Science, Anhui University, Hefei, Anhui Province 230601, People's Republic of China}

\begin{abstract}
We study amorphous systems with completely random sites
and find that, through constructing and exploring a concrete model Hamiltonian, such a system can host an exotic phase of topological amorphous metal in three dimensions. In contrast to the traditional Weyl semimetals, topological amorphous metals break translational symmetry, and thus cannot be characterized by the first Chern number defined based on the momentum space band structures. Instead, their topological properties will manifest in the Bott index and the Hall conductivity as well as the surface states. By studying the energy band and quantum transport properties, we find that topological amorphous metals exhibit a diffusive metal behavior. We further introduce a practical experimental proposal with electric circuits where the predicted phenomena can be observed using state-of-the-art technologies. Our results open a door for exploring topological gapless phenomena in amorphous systems.
\end{abstract}

\maketitle
Weyl semimetals, three-dimensional (3D) materials with Weyl points in band structures~\cite{Burkov2016,Jia2016,VishwanathRMP,XuReview}, have attracted considerable interest~\cite{Wan2011prb,Yuanming2011PRB,Burkov2011PRL,ZhongFang2011prl,Bergholtz2014PRL,
Tena2015RPL,Bergholtz2015PRL,Weng2015PRX,Shengyuan2015,Xie2015PRL,Xu2015,Lv2015,Lu2015,Zhangyi2016SciRep,Pixley2016PRX,Ueda2016,Yong2016typeii,Pixley2018PRL,Syzranov2018}
in recent years owing to their fundamental importance in mimicking Weyl fermions in particle physics and
their exotic topological properties. In the context of solid-state materials, the linear energy band dispersion around a Weyl point determines the semimetal
property with a zero density of states (DOS) at zero energy. In addition, the Weyl point is protected by the first Chern number defined by the integral of Berry curvatures over a closed surface in momentum space enclosing
the point~\cite{volovik}, leading to a Fermi arc consisting of surface states. This topological feature gives rise to the topological anomalous Hall effect~\cite{Yuanming2011PRB,Burkov2011PRL}. Beyond Weyl fermions that exist in particle physics, new fermions, such as type II Weyl fermions~\cite{Soluyanov2015Nature,Shuyun2016,Huang2016}
(also called structured Weyl fermions~\cite{Yong2015PRL}) and high spin fermions~\cite{Bernevig2016Science,HongDing2017Nature},
can appear in topological gapless materials.

All these topological gapless materials feature the existence of gapless structures in momentum space so that
the topological invariants can be further defined there. Yet, this can only be guaranteed in a crystalline material with
translational symmetry. Here, we ask whether a topological semimetal or metal can exist
in an amorphous system with completely random sites, such as glass materials, where the desired
translational symmetry is absent.
Recent development of technologies in engineering in quantum systems such as arbitrary positioning of
atoms~\cite{Browaeys2016Science,Lukin2016Science} and in mechanical systems such as constructing interacting gyroscopes
~\cite{MitchellNat2018} have paved the way for fabricating amorphous materials. However, the study of topological phenomena in amorphous systems is still in its infancy stage and only
a few works demonstrating the existence of topological insulators in amorphous systems has been reported~\cite{Shenoy2017PRL,MitchellNat2018,Ojanen2018NatCommun,Chong2017PRB,Xiao2017PRB,Prodan2018,Minarelli2018,Chern2018}. Whether topological
semimetals or metals exist in amorphous systems have not been explored hitherto.

\begin{figure}[t]
\includegraphics[width=2.8in]{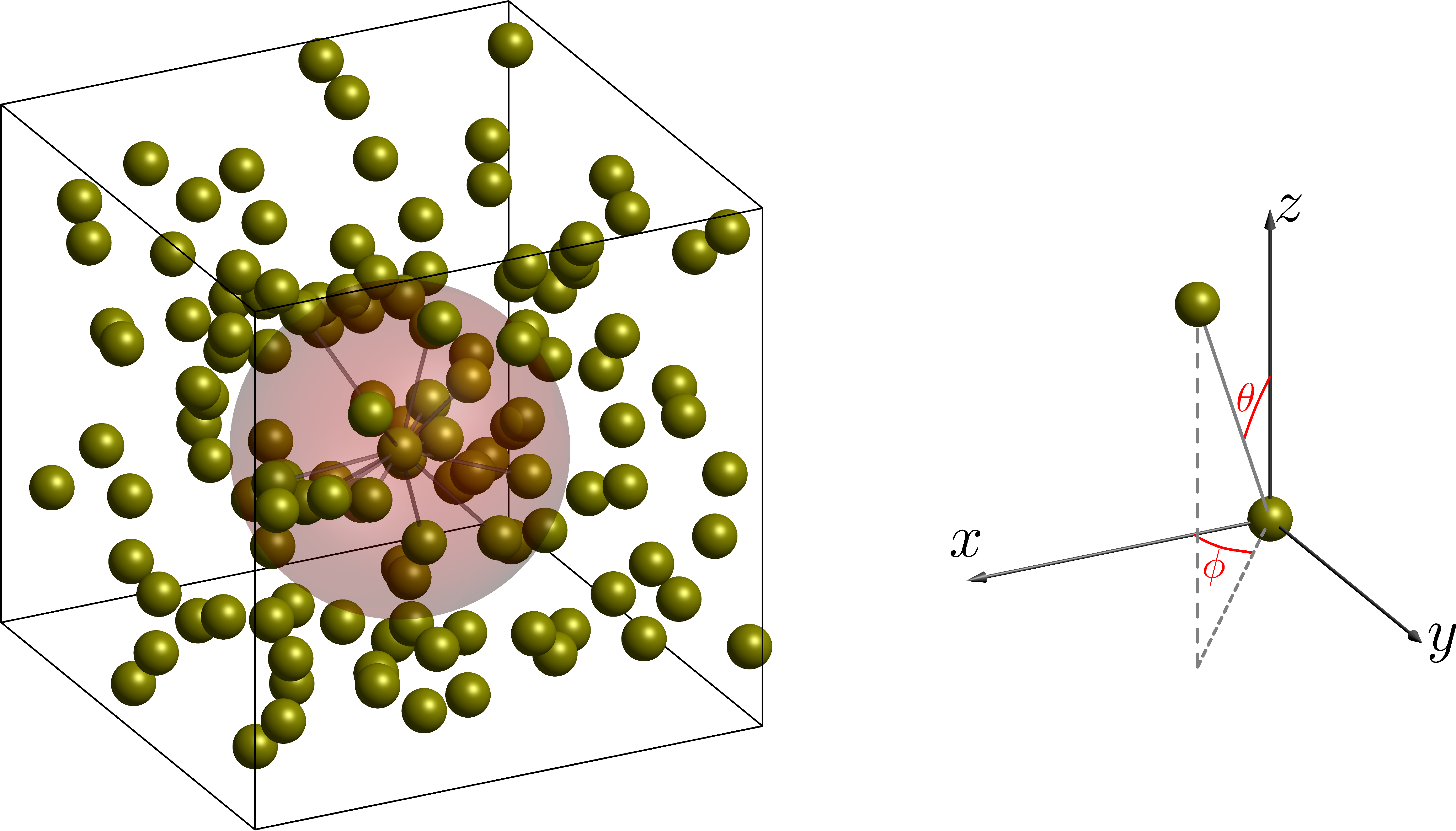}
\caption{(Color online) Schematic of a 3D random site configuration
with the allowed hopping inside the light red sphere for a typical site
at the center. }
\label{Fig1}
\end{figure}

In this paper, we demonstrate, by constructing and exploring a model Hamiltonian, the existence of a topological metal phase in a 3D amorphous system. The system is generated by randomly sampling sites in a box (see Fig.~\ref{Fig1}
for one sample configuration) and the results are
obtained by averaging over many sample configurations. We find three distinct phases, namely, the topological amorphous metal (TAM),
the amorphous Anderson insulator (AAI), and the amorphous insulator (AI) phases.
In contrast to Weyl semimetals with translational symmetry where their topology can be characterized by the first Chern number, the topological feature of our amorphous system is identified using the Bott
index, the Hall conductivity and the surface states.
To determine whether a phase in the amorphous system is a metal, a semimetal or an insulator, we compute the
band properties including the energy gap, the DOS, the level statistics and
the inverse participation ratio, and the transport properties including the longitudinal conductivity
and the Fano factor. We find that, in the most part of the parameter region where the Bott index and the Hall conductivity
are nonzero, the system is gapless, exhibiting a diffusive metal behavior. The other regions
correspond to the insulating phase where the longitudinal conductivity drops to zero and the Fano factor
suddenly rises to one. The insulator can be further divided into the AAI with
a nonzero DOS and the AI with a zero DOS. Finally, we introduce a practical  scheme to realize such a
Hamiltonian and observe its related exotic phenomena in electric circuits.

\emph{Model Hamiltonian.}--- We start by constructing the following model Hamiltonian
\begin{equation}
H=\sum_{\bf x}[\sum_{\bf R}t(R)\hat{c}_{\bf x}^\dagger H_0(\theta,\phi) \hat{c}_{{\bf x}+{\bf R}(\theta,\phi)} +
m_z\hat{c}_{\bf x}^\dagger\sigma_z \hat{c}_{\bf x}],
\end{equation}
where $\hat{c}_{\bf x}^\dagger=(\hat{c}_{{\bf x},\uparrow}^\dagger,\hat{c}_{{\bf x},\downarrow}^\dagger)$ with $\hat{c}_{{\bf x},\sigma}^\dagger$ creating a fermion with spin $\sigma$  at the position $\bf x$, which is a random vector
uniformly distributed
in the box, $x_\nu\in (0,L_\nu)$ with $\nu=x,y,z$, ${\bf R}(\theta,\phi)$ denotes the neighboring sites as shown in Fig.~\ref{Fig1}, $\sigma_\nu$ ($\nu=x,y,z$) are the Pauli matrices and $m_z$ is the mass term.
$H_0(\theta,\phi)=\sigma_z+i\sin\theta\cos\phi \sigma_x+i\sin\theta\sin\phi \sigma_y$
describes the hopping matrix for the neighboring sites as shown in Fig.~\ref{Fig1}.
We are inspired to construct such a Hamiltonian by the fact that it reduces to a well-studied Weyl semimetal model~\cite{VishwanathRMP}
when only the nearest-neighbor hopping is considered.
In light of irregular sites, we consider a case where the hopping strength decays exponentially
$t(R)=-e^{\lambda(1-R)}/2$, with $R$ being the spatial distance between two sites, where
we have chosen the units of energy and length to be one for simplicity. Here, we take
$\lambda=3$, the cutoff distance $R_c=2.5$ so that the hopping is neglected when $R>R_c$ \cite{Supplement}, and
the site density $\rho=N/V=1$ where $N$ is the number of sites and $V=L_x L_y L_z$ is the volume of the system.
For randomly distributed sets of $\bf x$, the system does not respect translational, time-reversal or inversion symmetries. Due to the random character, for numerical calculation, all our
results are averaged over 180-600 sample configurations.

\begin{figure}[t]
\includegraphics[width=3.2in]{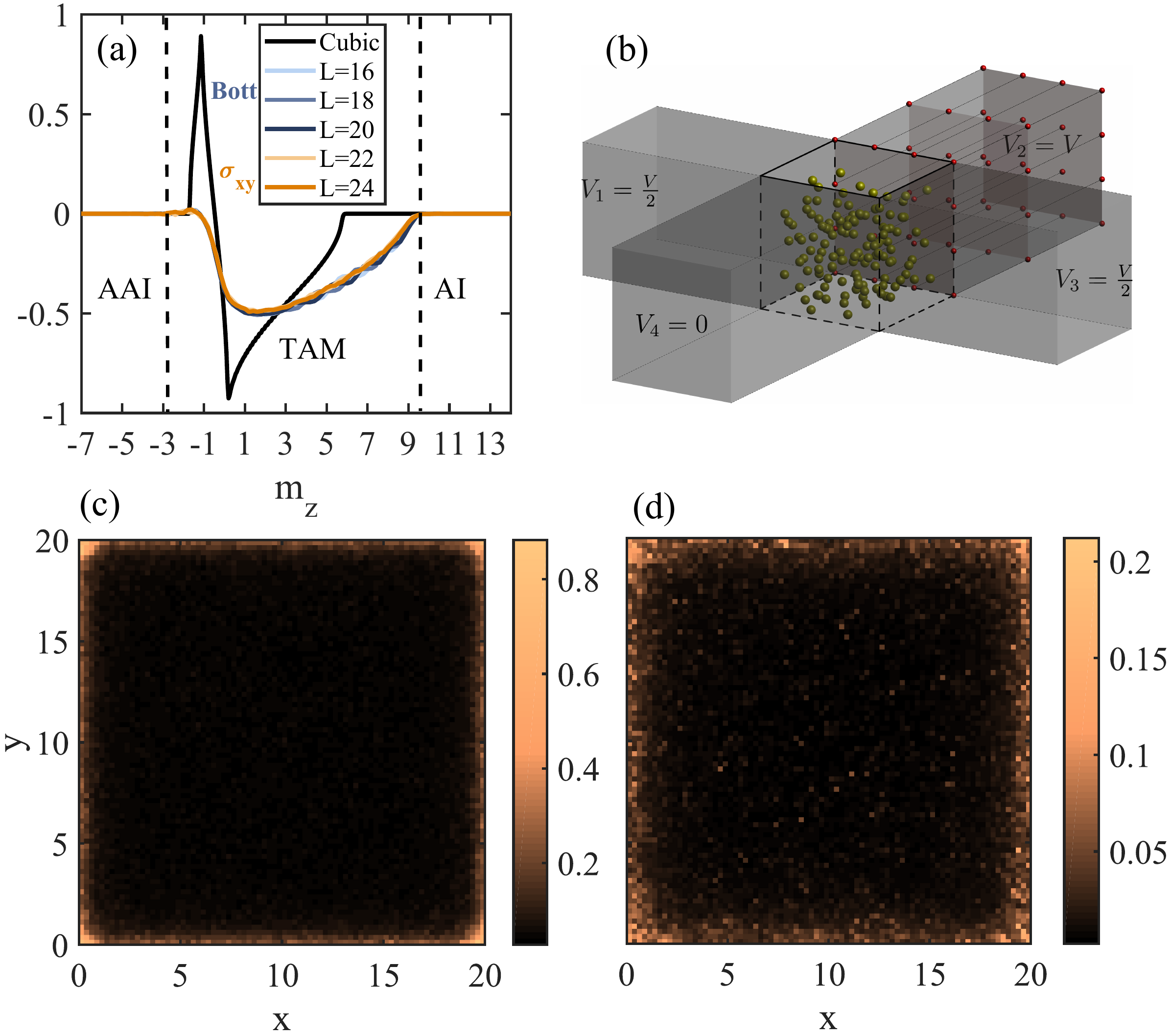}
\caption{(Color online) (a) The Bott index and the Hall conductivity in unit of $e^2/(2h)$
as a function of $m_z$ for distinct system sizes. The black line denotes the Bott index
for a cubic lattice configuration. Three different phases are identified:
amorphous Anderson insulator (AAI), topological amorphous metal (TAM) and amorphous
insulator (AI). (b) Schematic of a four terminal
setup used to compute the Hall conductivity, where we consider the cubic geometry
for all leads (see the dotted part for $V_2=V$). (c-d) The local density of states for
$m_z=2$ and $m_z=6$, respectively.}
\label{Fig2}
\end{figure}

In Fig.~\ref{Fig2}(a), we map out the phase diagram with respect to the mass $m_z$ incorporating three distinct phases (assuming that the Fermi surface lies at zero energy):
the TAM, the AAI and the AI phases,  according to the Bott index (or Hall conductivity) and the band and transport properties, which will be discussed
in detail in the following. For a topological phase, the Bott index is nonzero. For a diffusive metal,
the energy gap is zero, the DOS and conductivity are nonzero, and the Fano factor is 1/3.
For an insulator, the conductivity is zero and the Fano factor is 1. In our system, there are two types of
insulators: the Anderson insulator with a nonzero DOS and the band insulator with a zero DOS. Our results are
summarized in Table~\ref{tab1}.
\begin{table}[h]
\caption{Topological, band and transport properties of three distinct phases. Note
that, in the AI phase, the states around the zero energy are localized with
$\text{LSR}\sim 0.39$ and $\text{IPR}>0$.}
\label{tab1}
\setlength{\tabcolsep}{3pt}
\begin{tabular}{c  c  c  c  c  c  c  c}
\toprule
Phase & |Bott|($|\sigma_{xy}|$) & $\rho(0)$ & gap & $|\sigma_{zz}|$ & Fano factor & LSR & IPR\tabularnewline
\midrule
TAM & $>0$ & $>0$ & $\sim0$ & $>0$ & $\sim1/3$ & $\sim0.6$ & $\sim0$\tabularnewline
\midrule
AAI & $\sim0$ & $>0$ & $\sim0$ & $\sim0$ & $\sim1$ & $\sim0.39$ & $>0$\tabularnewline
\midrule
AI & $\sim0$ & $\sim0$ & $>0$ & $\sim0$ & $\sim1$ & --- & --- \tabularnewline
\bottomrule
\end{tabular}

\end{table}

\emph{Bott index and Hall conductivity.}--- In order to characterize the topology of the 3D amorphous system, we generalize the Bott index originally defined in 2D~\cite{firstPropBott} by defining it as
\begin{equation}
\mathrm{Bott}=\frac{1}{2\pi L_z}\text{Im}\text{Tr}\log(\tilde{U}_y\tilde{U}_x\tilde{U}_y^\dagger \tilde{U}_x^\dagger),
\end{equation}
where $\tilde{U}_x$ and $\tilde{U}_y$ are the reduced matrices for $U_x=\hat{P}e^{2\pi i\hat{x}/L_x}\hat{P}$ and $U_y=\hat{P}e^{2\pi i\hat{y}/L_y}\hat{P}$ in the occupied space, respectively.
Here, $\hat{x}$ and $\hat{y}$ are the position operators and $\hat{P}$ is the projection operator for the occupied space.
As we are interested in the case that the Fermi energy lies at zero energy,
we consider the states with negative energy as the occupied space for calculating the Bott index.
We prove that this generalized Bott index is equivalent to the topological anomalous Hall conductivity for a Weyl semimetal
(which is not necessary to be quantized) in Ref.\cite{Supplement}.

In Fig.~\ref{Fig2}(a), we plot the Bott index as a function of $m_z$ for different system sizes. Remarkably,
the amorphous system exhibits nonzero values for the Bott index when $-2.8\lesssim m_z\lesssim 9.6$,
suggesting the topological feature of the system. Compared with the cubic lattice configuration,
there appears a topologically nontrivial region for the amorphous system, which is topologically
trivial in a crystalline one. We can also observe that the absolute value of the Bott index is no longer symmetric with respect to $m_z$~\cite{Yanbin2018PRB} when the long-range hopping is involved; this explains why there only exists
a very small region with the positive Bott index. In addition,
the Bott index in the TAM region exhibits several plateaus, whose location
changes with respect to the system size, reflecting the finite size effect, similar to
the case of a crystalline Weyl semimetal.

To show that the Bott index reflects the Hall conductivity in a randomized system, we numerically calculate the
Hall conductivity using the Landauer-B\"uttiker formula in a mesoscopic system. We consider four ideal leads connected
to the amorphous system in the x and y directions as schematically shown in Fig.~\ref{Fig2}(b),
as we expect that a surface state appears on the surfaces vertical to these directions. Under the voltage $V_1=V_3=V/2$,
$V_2=V$ and $V_4=0$, the Hall conductivity is given by~\cite{DattaBook}
\begin{equation}
\sigma_{xy}=\frac{e^2}{2 hL_z}(T_{32}-T_{34}),
\end{equation}
where $T_{mn}$ is the total transmission probability from lead $n$ to $m$, which is computed using the nonequilibrium Green's function method~\cite{DattaBook,Sun2007PRB}. As $T_{32}-T_{34}$ accounts for the contribution from chiral edge modes,
for a Weyl semimetal, $\sigma_{xy}$ is equivalent to the Bott index multiplied by $e^2/(2h)$ and we
expect that this equivalence
also holds in an amorphous system.

\begin{figure}[t]
\includegraphics[width=3.4in]{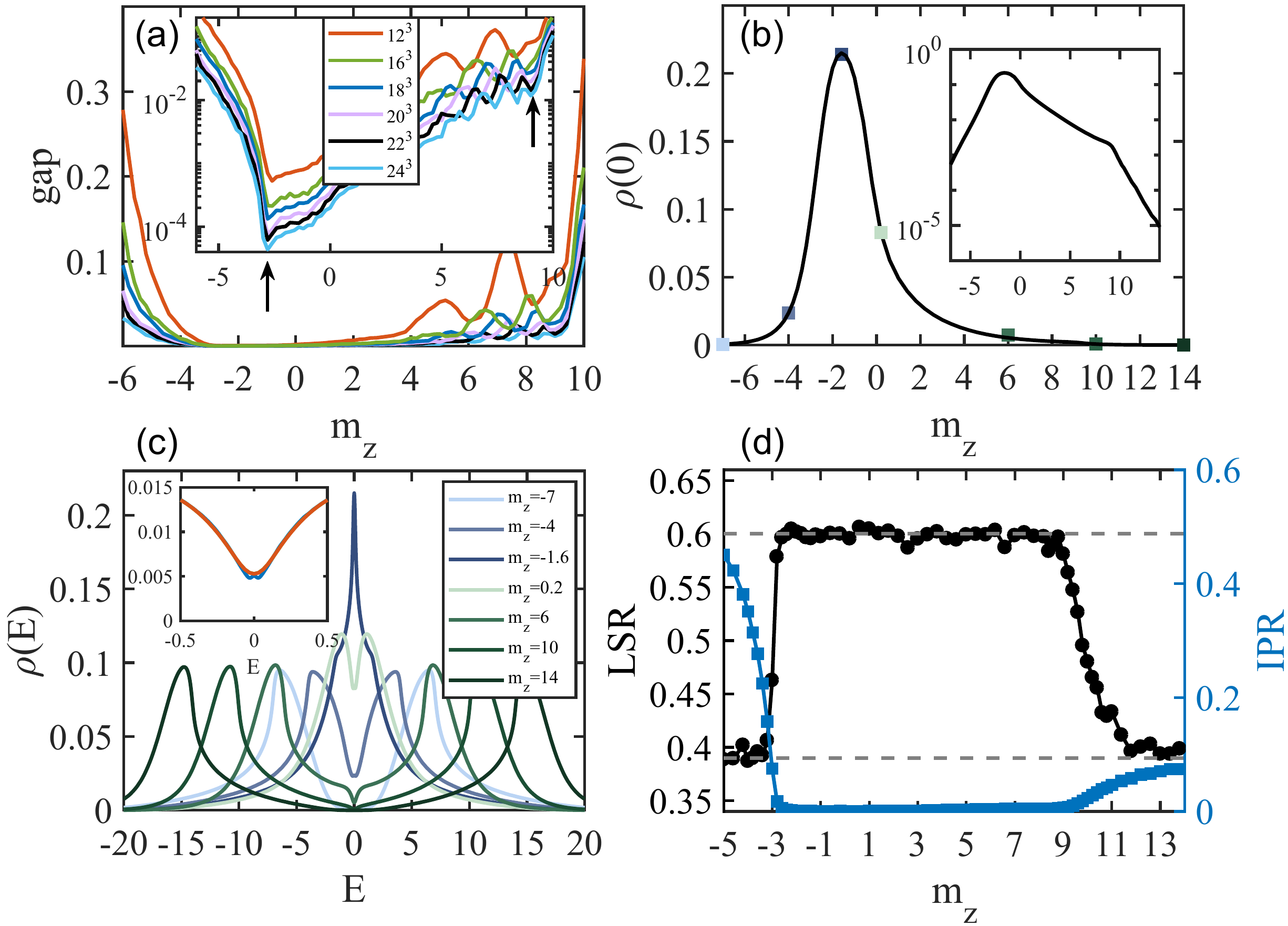}
\caption{(Color online) (a) The gap versus $m_z$ for different system sizes calculated via the
Lanczos method with the inset
plotting the same thing in the logarithmic scale. The arrows show the universal dips.
(b) The density of states (DOS) at zero energy $\rho(0)$ as a
function of $m_z$ (with the logarithmic scale figure shown in the inset) calculated by the kernel polynomial method (KPM) for $L=55$ and $N_c=2^{11}$. (c) The  DOS $\rho(E)$ versus $E$ for various $m_z$ in
different phases for $L=55$ and $N_c=2^{11}$. The inset plots $\rho(E)$ versus $E$ when $m_z=6$ for $L=55$, $N_c=2^{11}$
(red line), $L=60$, $N_c=2^{11}$ (green line), and $L=60$, $N_c=2^{13}$ (blue line). (d) The level spacing ratio (LSR) $r(E=0)$ (left vertical axis)
and  inverse participation ratio (IPR) $I(E=0)$ (right vertical axis) for $L=24$ for the states
around zero energy (only the states with negative energy are considered) computed via the Lanczos algorithm.
In all above figures,
samples in a cubic box are considered.}
\label{Fig3}
\end{figure}

In Fig.~\ref{Fig2}(a), we show the Hall conductivity in comparison to the Bott index.
We notice the clear consistence between them in a wide range of $m_z$ in an amorphous system as we expected.
For the slight discrepancy, we estimate that it is caused by finite size effects of the Bott index,
which exhibits conspicuous variations for distinct system sizes; the Hall conductivity does not show
clear finite size effects when $L=24$ as their difference from $L=22$ is small (we consider a cubic case, $L_x=L_y=L_z=L$). Further, the Hall conductivity does not exhibit clear plateaus from finite size effects probably due to the smearing out around the gap closing region as in Weyl semimetals. The nonzero Hall conductivity and Bott index suggest the existence of a topological amorphous phase in a wide range of parameters.

To further identify the topological feature of the system, we calculate the local DOS defined as
$
\rho(E,{\bf x})=[\sum_{i}\delta(E-E_i)(|\Psi_{E_i,\uparrow\bf x}|^2+|\Psi_{E_i,\downarrow\bf x}|^2)]$,
where $E_i$ is the $i$th eigenvalue, $\Psi_{E_i,\sigma\bf x}$ with $\sigma=\uparrow,\downarrow$ are the corresponding components of the
eigenstate of the system, and $[\cdots]$ denotes the average over samples. The DOS is defined as
$\rho(E)=\sum_{\bf x}\rho(E,{\bf x})/(2N)$, which is normalized to one, i.e., $\int dE\rho(E)=1$.
In Fig.~\ref{Fig2}(c) and (d), we illustrate the local DOS summed over $x_z$: $\sum_{x_z}\rho(E,{\bf x})$ for
a system $L_x=L_y=20$ and $L_z=10$ for two typical values of $m_z$, clearly showing the presence of
the surface states localized on the boundaries \cite{Supplement}.

\emph{Band properties.}--- To discriminate the metal or semimetal phase from the
insulator phase with respect to $m_z$, we compute the gap, twice of the lowest positive energy,
using the Lanczos algorithm, and the DOS for large systems using the kernel polynomial method (KPM), which
expands the DOS in Chebyshev polynomials to the order $N_c$~\cite{Alexander2006RMP}.

Figure~\ref{Fig3}(a) and (b) illustrate the gap and the DOS at zero energy $\rho(E=0)$ with respect to the mass $m_z$ for distinct system sizes. Clearly, we see that the region with nonzero Bott index from $-2.8\lesssim m_z\lesssim9$ corresponds to the gapless region:
The gap for $-3.2<m_z<2$ is very small even for a small system size (see
the red line for $L=16$) associated with a relative large DOS.
$\rho(E=0)$ reaches the maximum at $m_z=-1.6$, where $\rho(E)$ versus $E$ exhibits a steep peak at zero energy
as shown in Fig.~\ref{Fig3}(c), and it decreases sharply as $m_z$ moves away from this point associated with a
developed minimum around zero energy for $\rho(E)$ (see Fig.~\ref{Fig3}(c)).
When $2\lesssim m_z\lesssim9$,
while the energy gap strongly depends on the system size, its overall decline with increasing the system size can be observed,
suggesting that this phase may be a semimetal or metal. Figure~\ref{Fig3}(b) further shows that $\rho(E=0)$
does not vanish in this region despite being small, implying that they correspond to a metal phase instead of a semimetal one.
Specifically, for $m_z=6$, $\rho(E)$ shows a sudden drop around zero energy (see Fig.~\ref{Fig3}(c)), but this minimum does
not vanish. To exclude the finite size effect, we calculate $\rho(E)$ using larger system size and $N_c$ and do not find
conspicuous decline of $\rho(E=0)$ as shown in the inset of Fig.~\ref{Fig3}(c) \cite{Supplement}, in stark contrast to a dramatic drop in a Weyl semimetal with quasiperiodic disorder~\cite{Pixley2018PRL}.

Viewing Fig.~\ref{Fig3}(a) in the logarithmic system (see the inset), we clearly see that there appears
a universal dip of the energy gap for different system sizes at $m_z=9$ and $m_z=-2.8$.
For the former, $\rho(E=0)$ exhibits a rapid decline to zero as $m_z$ increases from this point (see the inset in Fig.~\ref{Fig3}(b)),
suggesting a phase transition to a band insulator
[see $\rho(E)$ versus $E$ for $m_z=10,14$ in Fig.~\ref{Fig3}(c)].
More interestingly, for the latter, the DOS does not vanish and does not show clear
nonanalytic behavior with respect to $m_z$. This phase is actually the Anderson localized
insulator (dubbed amorphous Anderson insulator), which will be identified by the
level-spacing statistics, the inverse participation ratio (IPR), the conductivity
and the Fano factor. We note that with the further decline of $m_z$, the system develops
into a band insulator [see $\rho(E)$ versus $E$ for $m_z=-7$ in Fig.~\ref{Fig3}(c)], but
we cannot identify the transition point since the DOS becomes very small.

For level statistics, we calculate the
adjacent level-spacing ratio (LSR):
$
r(E)=[\frac{1}{N_E-2}\sum_{i}\text{min}(\delta_i,\delta_{i+1})/\text{max}(\delta_i,\delta_{i+1})]$,
where $\delta_i=E_i-E_{i-1}$ with $E_i$ being the $i$th eigenenergy sorted in
an ascending order and $\sum_{i}$ is the sum over an energy bin around the energy $E$
with $N_E$ being the energy levels counted.
It is well known that for localized states, $r\approx 0.39$~\cite{Huse2007PRB} associated with the Poisson
statistics and for extended states, $r\approx 0.6$ corresponding to the Gaussian unitary
ensemble (GUE)~\cite{Rigol2014PRX}. Another
signature we use is the real space IPR:
$I(E)=[\frac{1}{N_E}\sum_i\sum_{\bf x}(|\Psi_{E_i,\uparrow\bf x}|^2+|\Psi_{E_i,\downarrow\bf x}|^2)^2]$,
which measures how much a state
around energy $E$ is spatially localized. For a completely extended state in an infinitely
large system, it is zero; for a state localized in a single site, it is one.

Figure~\ref{Fig3}(d) shows that, in the topological metal regime,
$r(E=0)$ is around $0.6$ and $I(E=0)$ is almost zero; when $m_z$ decreases from $-2.8$,
$r(E=0)$ drops to around $0.39$ and $I(E=0)$ increases sharply,
indicating the phase transition from the extended phase to the localized one.
Interestingly, we also see a similar change of the LSR and the IPR around $m_z\sim 9$, implying
that the states around zero energy are localized even though the DOS becomes very small \cite{Supplement}.

\emph{Conductivity and Fano factor.}--- To study the quantum transport properties of the amorphous system,
we numerically calculate the transmission matrix $tt^\dagger$ at zero energy by the nonequilibrium Green's function method~\cite{DattaBook,Sun2007PRB} and determine the zero-temperature conductance by the Landauer formula $G=(e^2/h)\text{Tr} (tt^\dagger)$~\cite{DattaBook} and the Fano factor $F=\text{Tr}[tt^\dagger(1-tt^\dagger)]/\text{Tr}(tt^\dagger)$~\cite{Beenakker1992PRB,Bergholtz2014PRL},
for a system connected to two ideal terminals for $z<0$ and $z>L_z$.

\begin{figure}[t]
\includegraphics[width=3.2in]{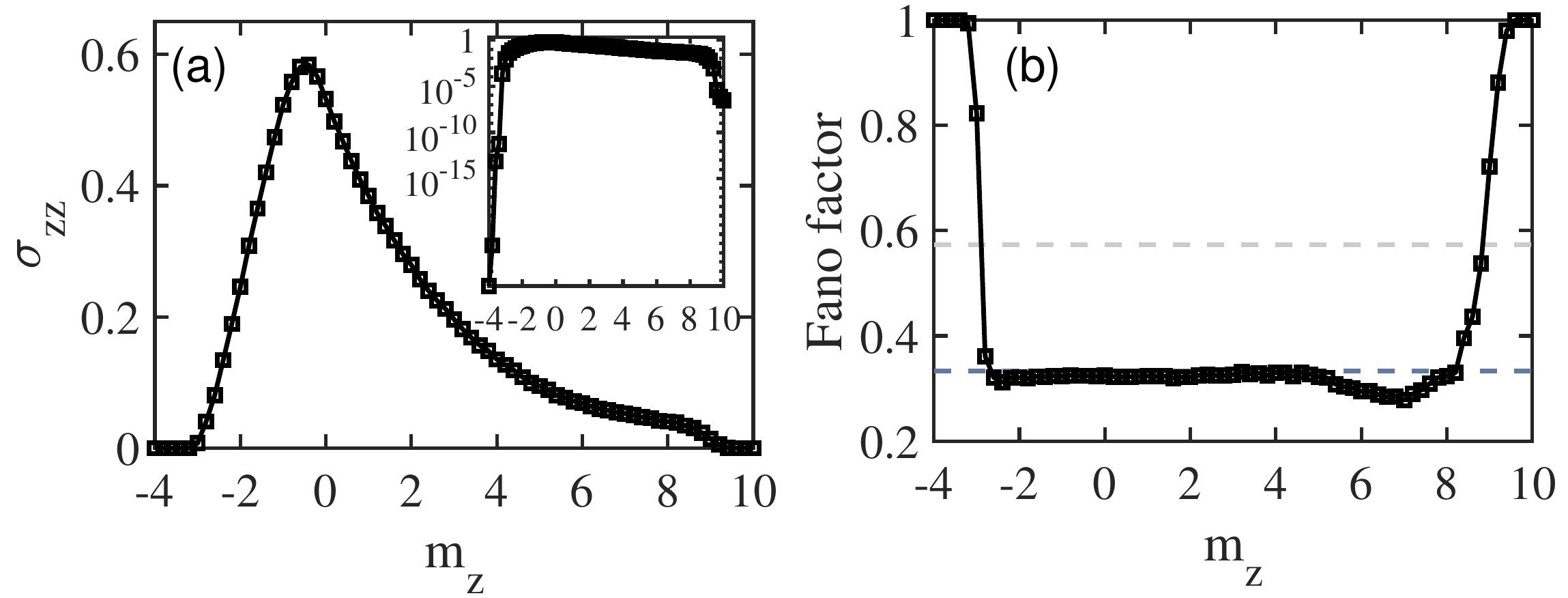}
\caption{(Color online) Conductivity $\sigma_{zz}$ (a) in unit of $e^2/h$ and Fano factor (b) versus $m_z$ for $L=25$ in a cubic box. The inset plots the conductivity
in the logarithmic scale, showing its steep drops across the phase transitions.
The dashed lines correspond to $F=1/3$ and $F=1/3+1/(6\ln 2)$, respectively.}
\label{Fig4}
\end{figure}

Figure~\ref{Fig4}(a) shows the conductivity $\sigma_{zz}=LG/W^2$ versus $m_z$ with $W$ and $L$ being the width and length of the system (we here consider a cubic box geometry, i.e., $W=L$ ).
The conductivity is nonzero in the region with nonzero Bott index from $-2.8\lesssim m_z\lesssim 9$,
showing a diffusive metal behavior as for a pseudoballistic semimetal the conductivity vanishes~\cite{Bergholtz2014PRL}.
The conductivity drops to zero at around $m_z\sim-2.8$ and at around $m_z\sim9$ when $m_z$ moves away
to the left and right region, respectively. The former corresponds to the transition into the
Anderson insulator phase, while the latter the band insulator phase with vanishing DOS.
The diffusive metal behavior is also reflected in the Fano factor that takes the value around $F=1/3$~\cite{Beenakker1992PRB}
(see Fig.~\ref{Fig4}(b)). The transition into the insulator phase is signalled by the steep rise
of the Fano factor to one due to the Poisson process. We do not find any discernible
region where the Fano factor takes the value of $F_0=1/3+1/(6\ln 2)$ for Weyl semimetals without disorder~\cite{Bergholtz2014PRL}, further suggesting the absence of the semimetal phase \cite{Supplement}.

\emph{Experimental realization.}--- Topological amorphous metals may be realized in classical systems, artificial quantum systems and solid-state glass materials.
Here, we propose an experimental scheme to engineer a Laplacian (acting as a Hamiltonian) with electric circuits, which takes the form of our Hamiltonian \cite{Supplement}. The surface states can be observed by measuring the two-point impedance. Recently, a number of topological phases, such as the SSH model~\cite{Thomale2018CP}, Weyl semimetals~\cite{Simon2019PRB} and higher topological
insulators~\cite{Thomale2018NP} have been experimentally observed with electric circuits. In addition, recent development of technology has allowed us to place Rydberg atoms in arbitrary geometry using optical tweezers~\cite{Browaeys2016Science,Lukin2016Science},
which makes it possible to realize our model in this system.

In summary, we have discovered a topological amorphous metal phase in
3D amorphous systems. We identify its topological feature by calculating the Bott
index, the Hall conductivity and the surface states. Through further studying its band properties including
the energy gap, DOS, LSR and IPR and the quantum transport properties, we find that
the topological phase exhibits a diffusive metal behavior. We further predict the
phase transition from the topological metal phase to the Anderson insulator phase
and the band insulator phase with respect to a system parameter.
Our results open a new avenue for studying topological gapless phenomena in amorphous systems.
These new phenomena might be observed in various amorphous materials, such as
engineered classical or atomic systems and glass materials.

\begin{acknowledgments}
We thank A. Agarwala, H. Jiang, S.-G. Cheng and H.-W. Liu for helpful discussions. Y.B.Y. and Y.X.
are supported by the start-up fund from Tsinghua University (53330300219) and
the National Thousand-Young-Talents Program (042003003). T.Q. is supported by the start-up fund
(No.S020118002/069) from Anhui University. D.L.D. acknowledges the start-up fund from Tsinghua University.
L.M.D. is supported by the Ministry of Education and the National Key Research and
Development Program of China (2016YFA0301902).
\end{acknowledgments}

\begin{widetext}
\section{Supplemental Material}
\setcounter{equation}{0} \setcounter{figure}{0} \setcounter{table}{0} %
\renewcommand{\theequation}{S\arabic{equation}} \renewcommand{\thefigure}{S%
\arabic{figure}} \renewcommand{\bibnumfmt}[1]{[S#1]} \renewcommand{%
\citenumfont}[1]{S#1}

In the supplementary material, we will prove the equivalence between the generalized Bott index
and the Hall conductivity in 3D Weyl semimetals in Section 1, show the Griffiths effects in
Section 2 and the density profiles of the surface states in Section 3, give more discussion
on the semimetal phase in Section 4, illustrate the mobility edge in distinct phases in Section 5 and
the effects of $R_c$ and the on-site disorder in Section 6, and finally introduce an
experimental scheme with electric circuits in Section 7.

\section{S-1. Proof for the equivalence between the Bott index and the Hall conductivity}
In this section, we will prove the equivalence between the
Bott index defined in the main text and the Hall conductivity in 3D Weyl semimetals
with translational symmetry, following the method used to prove its equivalence to the
Chern number in 2D systems~\cite{Rigol2017PRA}. For a Weyl semimetal, let us define the Bott index as
\begin{equation}
\mathrm{Bott}_3=\frac{1}{2\pi L_3}\text{Im}\text{Tr}\log(U),
\end{equation}
where $U=\tilde{U}_2\tilde{U}_1\tilde{U}_2^\dagger \tilde{U}_1^\dagger$, $U_i=Pe^{2\pi i \hat{{\bf r}} \cdot {\bf a}_i/(L_i a_i)}P=\left(
                                                                  \begin{array}{cc}
                                                                    0 & 0 \\
                                                                    0 & \tilde{U}_i \\
                                                                  \end{array}
                                                                \right)
$ with the position operator $\hat{\bf r}=\sum_{i=1,2,3}\hat{x}_i{\bf a}_i$, ${\bf a}_i$ being the
lattice vectors and $L_i$ being the size of the system along the ${\bf a}_i$ direction;
$\tilde{U}_i$ is the reduced matrix in the occupied space and
$P$ is the projection operator that projects states into the occupied space. In a system with translational symmetry,
$P$ can be expressed as $P=\sum_{n{\bf k}}|n{\bf k}\rangle\langle n {\bf k}|$ where $|n{\bf k}\rangle$
denotes the occupied Bloch state in the $n$th band with the quasimomentum ${\bf k}=\sum_{i=1,2,3}k_i {\bf G}_i/(2\pi)$,
where ${\bf G}_i$ is the reciprocal lattice vector.
In the coordinate representation, the Bloch state takes the form of $\langle {\bf r}|n,{\bf k}\rangle=e^{i{\bf k}\cdot {\bf r}}u_{n,{\bf k}}({\bf r})
=e^{i{\bf k}\cdot{\bf r}}\langle {\bf r}|u_{n,{\bf k}}\rangle$ where $u_{n,{\bf k}}({\bf r+{\bf a}_i})=u_{n,{\bf k}}({\bf r})$.
In this basis, we can expand $U$ in terms of $\delta k_i$ with $\delta k_i=2\pi a_i/L_i$ ($i=1,2$),
\begin{eqnarray}
U=&&\sum_{n_1,n_2,...,n_5}\sum_{{\bf k}} |n_1,{\bf k}\rangle
\langle u_{n_1,{\bf k}}|u_{n_2,k_1,k_2-\delta k_2,k_3}\rangle
\langle u_{n_2,k_1,k_2-\delta k_2,k_3}|u_{n_3,k_1-\delta k_1,k_2-\delta k_2,k_3}\rangle \nonumber \\
&&\langle u_{n_3,k_1-\delta k_1,k_2-\delta k_2,k_3}| u_{n_4,k_1-\delta k_1,k_2,k_3} \rangle
\langle u_{n_4,k_1-\delta k_1,k_2,k_3}|u_{n_5,{\bf k}}\rangle \langle n_5,{\bf k}| \\
=&&U_0+U_2+O(\delta k^3),
\end{eqnarray}
where
\begin{eqnarray}
U_0&=&\sum_{n}\sum_{\bf k}|n\rangle \langle n|, \\
U_2&=&\sum_{n_1,n_2}\sum_{\bf k}
|n_1\rangle \langle n_2| \big[ \delta k_1 \delta k_2 \big( (\partial_{k_2}\langle u_{n_1}|)(\partial_{k_1} |u_{n_2}\rangle)-k_1\leftrightarrow k_2 \big) \big. \nonumber \\
&&\big.+
\frac{1}{2}\sum_{i=1,2}\delta k_i^2 (\langle u_{n_1}|\partial^2_{k_i}
u_{n_2}\rangle + \langle \partial^2_{k_i} u_{n_1}|u_{n_2}\rangle) \big] \nonumber \\
&&+ \sum_{n_1,n_2,n_3}\sum_{\bf k} |n_1\rangle \langle n_3|
\big[ \sum_{i=1,2}\delta k_i^2 \langle u_{n_1}|\partial_{k_i} u_{n_2}\rangle\langle \partial_{k_i} u_{n_2}|u_{n_3}\rangle \nonumber \\
&&+\delta k_1 \delta k_2 \langle u_{n_1}|\partial_{k_1}u_{n_2}\rangle\langle \partial_{k_2} u_{n_2}|u_{n_3}\rangle
+\delta k_1 \delta k_2 \langle \partial_{k_2} u_{n_1}|u_{n_2}\rangle\langle \partial_{k_1} u_{n_2}|u_{n_3}\rangle \big],
\end{eqnarray}
where $n_i$ denotes the occupied bands, and, for briefness, we have skipped the
index for ${\bf k}$. Let us further decompose $U$ into
$U=1+U_{\bar{D}}+U_{O}$ where $U_{\bar{D}}+1$ denotes the diagonal part and
$U_{O}$ the off-diagonal one. Using this decomposition, we obtain
\begin{equation}
\text{Tr}\log U=\text{Tr}U_{\bar{D}}-\frac{1}{2}\text{Tr}(U_{\bar{D}}+U_O)^2+O((U_{\bar{D}}+U_O)^3).
\end{equation}
Since $U_{\bar{D}}+U_O$ is at least the second order of $\delta k$, the second term contributes a fourth order term
and hence we neglect it. To the second order, we only need to evaluate the diagonal part, which is
\begin{eqnarray}
\text{Tr}U_{\bar{D}}=\text{Tr}U_2
&=&\sum_{n}\sum_{\bf k}\big[ \delta k_1 \delta k_2 \big( \langle \partial_{k_2}u_{n}|\partial_{k_1}u_{n}\rangle-k_1\leftrightarrow k_2 \big)
+
\frac{1}{2}(\sum_{i=1,2}\delta k_i^2 \langle u_{n}|\partial^2_{k_i}|
u_{n}\rangle +c.c. ) \big] \nonumber \\
&&+ \sum_{n}\sum_{\bf k}\sum_{n^\prime}
\sum_{i=1,2}\delta k_i^2 |\langle u_{n}|\partial_{k_i} u_{n^\prime}\rangle|^2,
\end{eqnarray}
where only the first term contributes to the Bott index as all other terms are purely real.
Therefore, to the second order, we have
\begin{eqnarray}
\mathrm{Bott}_3&=&\frac{1}{2\pi L_3} \sum_{n}\sum_{\bf k}\delta k_1 \delta k_2 \Omega_3({\bf k}) \\
&=&\frac{1}{2\pi}\sum_n \int_0^{2\pi} dk_3 C_n(k_3),
\end{eqnarray}
where $\Omega_3({\bf k})=i(\langle \partial_{k_1}u_{n}|\partial_{k_2}u_{n}\rangle-k_1\leftrightarrow k_2)$ is
the Berry curvature along the ${\bf G}_3$ direction and $C_n$ is the Chern number for a fixed $k_3$ in the
$n$th occupied band. Evidently, this is the Hall conductivity $\sigma_{12}$ in unit of $e^2/h$ and hence we prove the equivalence
between the Bott index and the Hall conductivity in a Weyl semimetal.

\begin{figure}[t]
\includegraphics[width=4in]{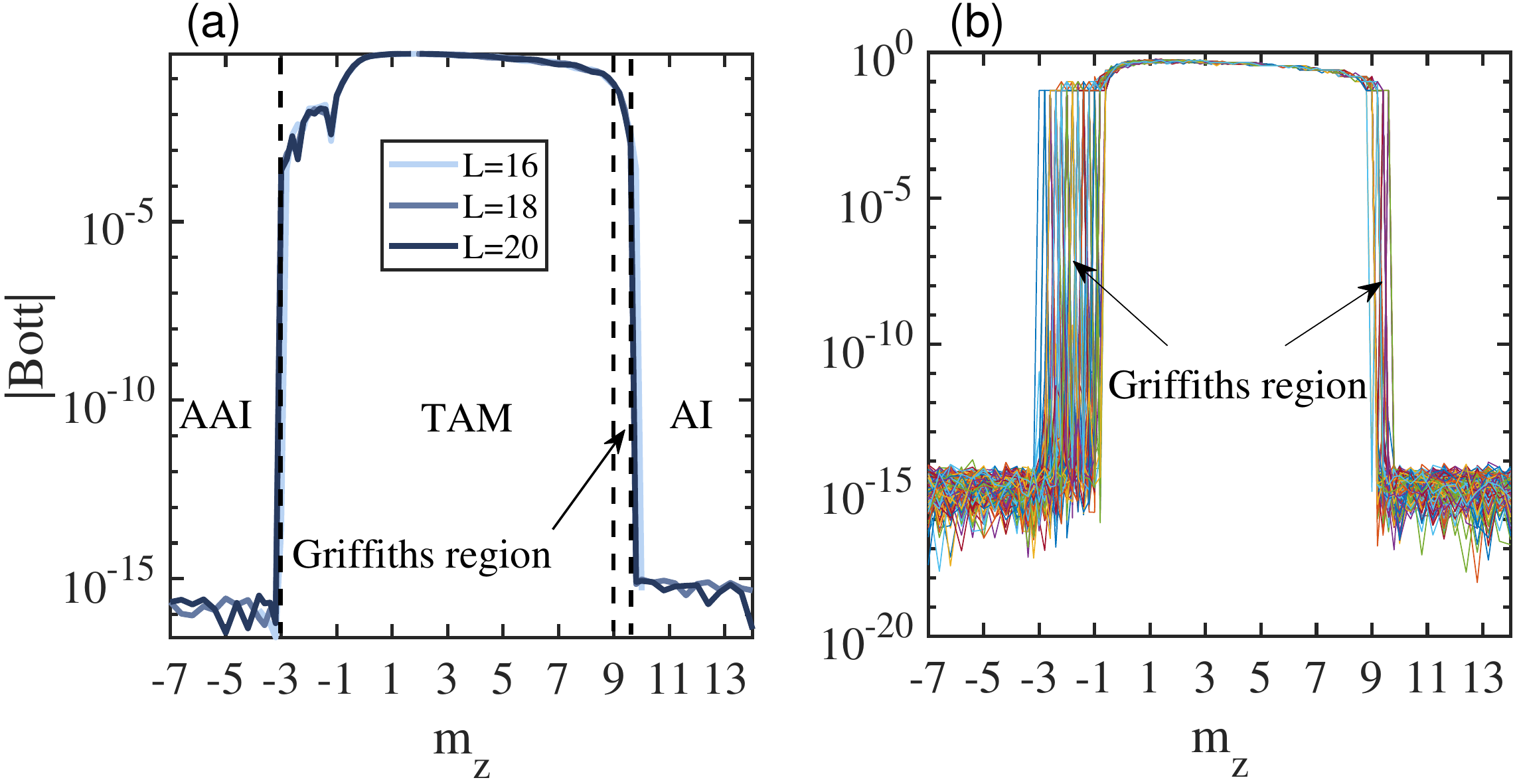}
\caption{(Color online) The absolute value of the averaged Bott index (a) and the Bott index for
181 samples for $L=20$ (b) with respect to $m_z$ in the logarithmic scale. }
\label{SMFig1}
\end{figure}

\begin{figure}[t]
\includegraphics[width=4in]{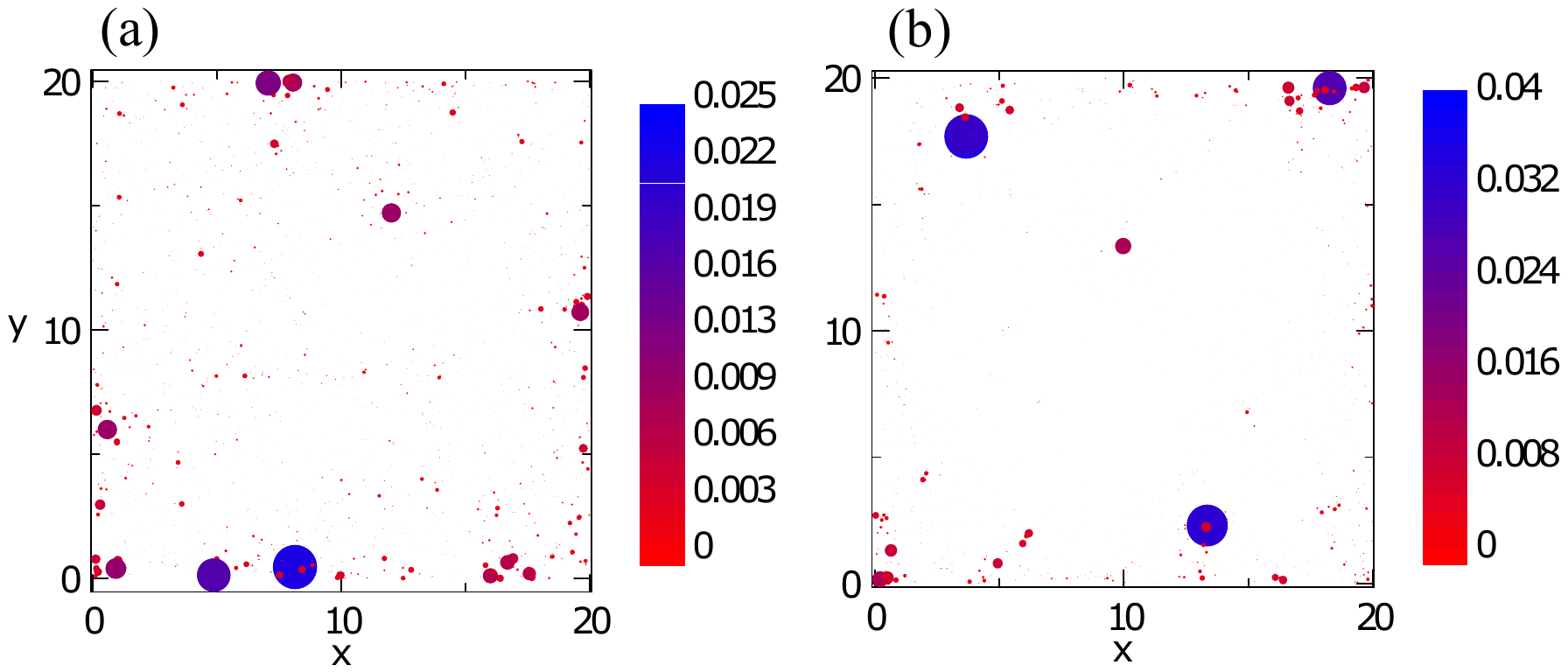}
\caption{(Color online) Top view of the surface states
for two typical states for $m_z=2$ (a) and $m_z=6$ (b). Here, the color and the size of circles depict the profile of the surface states, which are clearly localized around the boundaries. }
\label{FigSM2}
\end{figure}

\section{S-2. Griffiths region}
In Fig.~\ref{SMFig1}, we plot the absolute value of the Bott index
in the logarithmic scale, which clearly shows that the Bott index drops to zero across the phase boundaries.
We have also noticed that, in the region $9<m_z<9.6$, while the system transitions into an
insulating phase, the Bott index does not vanish despite being small. To interpret such phenomena,
we plot the absolute value of the Bott index for all 181 samples in Fig.~\ref{SMFig1}(b), illustrating that, in this region, some samples
have nonzero Bott index while others have zero, suggesting that this region corresponds to a Griffiths region.

\section{S-3. Surface states}
In the main text, we show the local DOS in a flat-box like geometry with
the height much shorter than the other two dimensions (here we take $L_x=L_y=20$ and $L_z=10$).
Here, we use the same geometry so that
the system is gapped under periodic boundary conditions. This enables us to pick up the surface
states that are located inside the gap under open boundary conditions along the x and y directions.
We illustrate the top view of the surface states in Fig.~\ref{FigSM2}(a) and (b), clearly showing
their localization on the boundaries. The IPR of the two states are 0.0027 and 0.0066, respectively,
which are in the same order of $1/(4L_x L_z)=0.0013$, the IPR for a state uniformly distributed on
the surfaces.

\section{S-4. Discussion on semimetal phases}
\begin{figure}[t]
\includegraphics[width=4in]{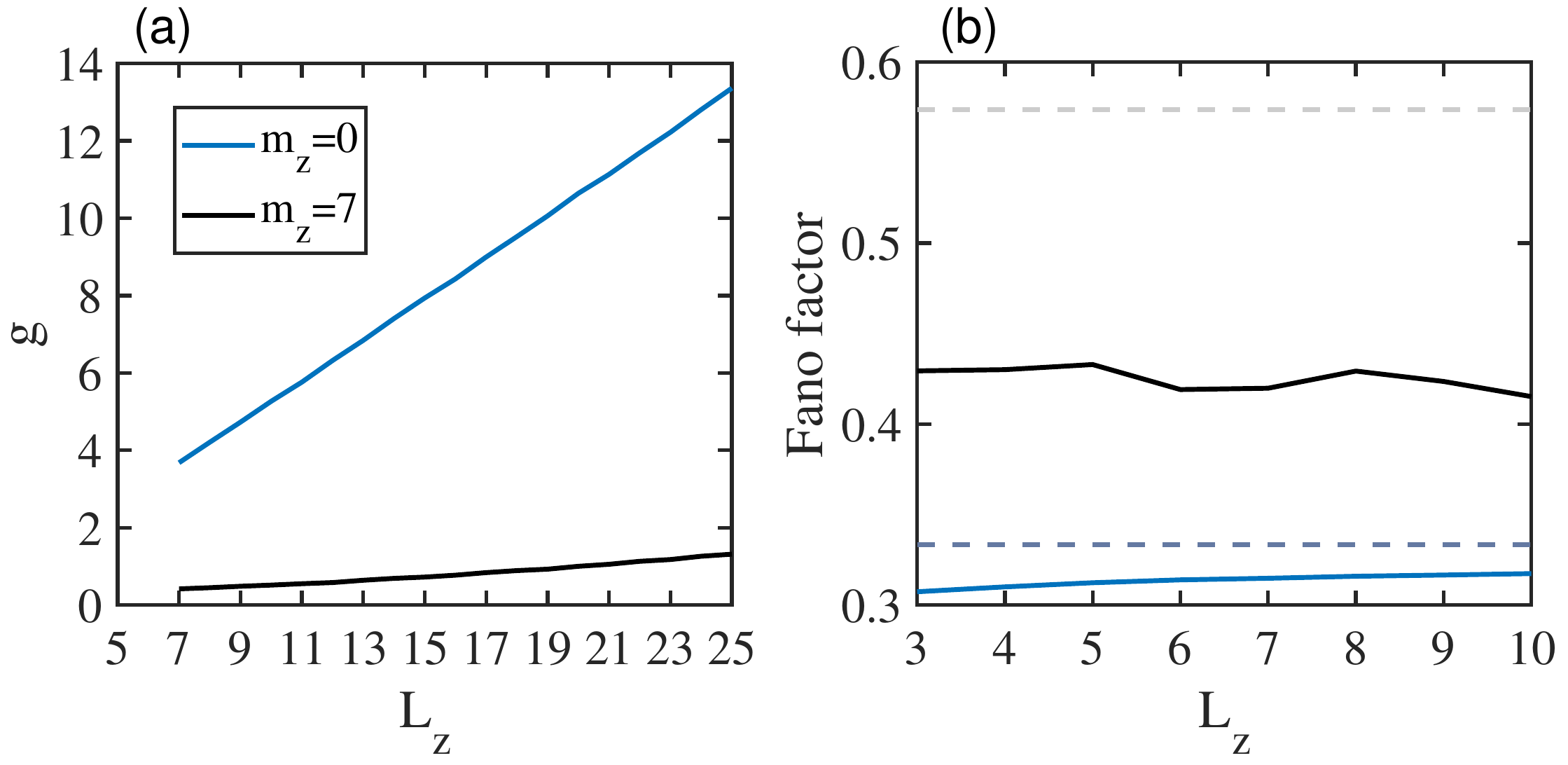}
\caption{(Color online) (a) $g$ versus $L_z$ for $L_x=L_y=25$ and (b) the Fano factor versus $L_z$
for $L_x=L_y=40$.}
\label{FigSM3}
\end{figure}

To further verify the absence of the semimetal phase, let us study the intrinsic conductivity
and the Fano factor in a flat-box geometry. By defining a dimensionless parameter
\begin{equation}
g=G\frac{h}{e^2}\frac{L_z^2}{W^2},
\end{equation}
we write the intrinsic conductivity (eliminating the contact resistance contribution) as
\begin{equation}
\sigma_I=\frac{1}{W^2\partial R/\partial L_z}=\frac{e^2}{h}\frac{1}{\partial (L_z^2/g)/\partial L_z},
\end{equation}
where $R=1/G=\frac{L_z^2}{W^2 g} \frac{h}{e^2}$ is the resistance. For a diffusive metal, $g\propto L_z$ as $L_z\rightarrow \infty$
so that $\sigma_I$ is finite, while in a pseudo-ballistic regime corresponding to a semimetal, $g$ has
an upper bound as $L_z\rightarrow \infty$ so that $\sigma_I$ goes zero. In Fig.~\ref{FigSM3}(a), we see
that for both $m_z=0$ and $m_z=7$, $g$ increases with $L_z$ with a linear scaling, suggesting that
the $\sigma_I$ do not vanish in both cases, while the slope is much smaller for $m_z=7$.
In Fig.~\ref{FigSM3}(b), we further plot the Fano factor using a flat-box like geometry with $W=40$ as a
function of $L_z$, illustrating that the Fano factor is slightly below $1/3$ for $m_z=0$ and increases slowly with
increasing $L_z$, while for $m_z=7$, the value stays around $0.43$, which is between the value (0.5738) for a semimetal
and $1/3$ for a metal. As the geometry becomes a cubic, the Fano factor for $m_z=0$ goes to $1/3$ while for $m_z=7$
goes below $1/3$ (see Fig. 4(b) in the main text).

\section{S-5. Mobility edges}
\begin{figure}[t]
\includegraphics[width=5in]{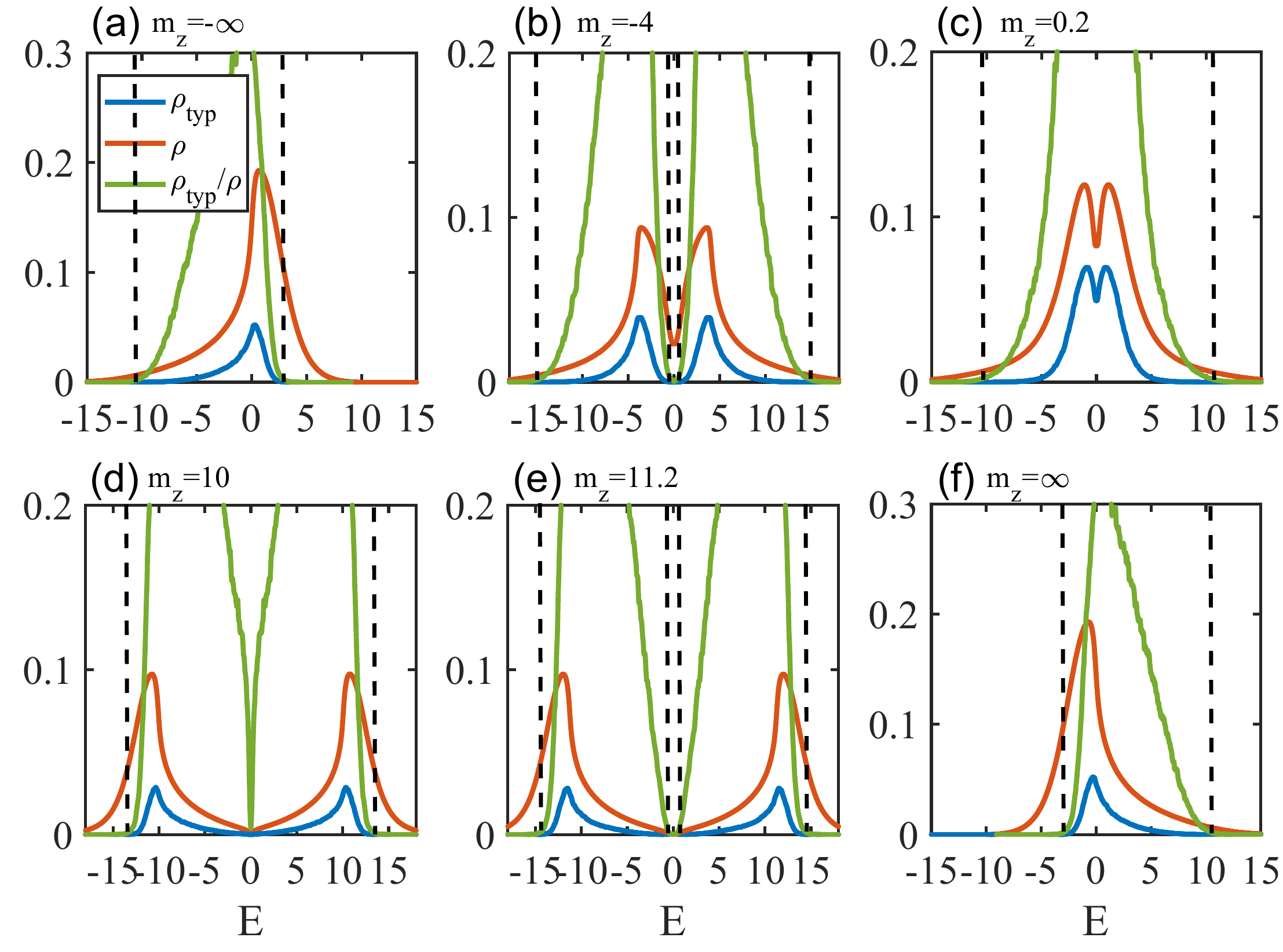}
\caption{(Color online) The typical DOS, the DOS and their ratio versus $E$ for different values of $m_z$.
In (a) and (f) where $|m_z|\rightarrow \infty$, the constant energy $m_z$ is not included
and only the lower band information is plotted.
The dashed black lines denote the mobility edge.
Here, the DOS and the typical DOS are numerically calculated with $L=55$, $N_c=2^{11}$,
and $L=55$, $N_c=2^{12}$ in a cubic box, respectively.}
\label{FigSM4}
\end{figure}

In this section, we study the mobility edge in our system where the system transitions into the extended phase from the localized one as the Fermi surface is tuned.
For clarity, let us first consider the limit that $m_z\rightarrow -\infty$. Based on the perturbation theory to the first order, we obtain
the following effective Hamiltonian
\begin{equation}
H_{eff}=\sum_{\bf x}\left[\sum_{\bf R}t(R)\hat{c}^\dagger_{{\bf x},\uparrow}\hat{c}_{{\bf x}+{\bf R},\uparrow}+
m_z\hat{c}^\dagger_{{\bf x},\uparrow}\hat{c}_{{\bf x},\uparrow}\right],
\end{equation}
which describes a spinless particle in a 3D amorphous system. In Fig.~\ref{FigSM4}(a), we plot the DOS of this
system without including the constant term $m_z$. The DOS is asymmetric with respect to $E$. For a cubic lattice
configuration, it shares the asymmetric characteristic due to the presence of the long-range hopping, in
stark contrast to the case with only the nearest-neighbor hopping.

To determine the mobility edge, we calculate the typical DOS defined as
\begin{equation}
\rho_{typ}(E)=e^{[\frac{1}{N}\sum_{{\bf x}}\text{log}\rho(E,{\bf x})]},
\end{equation}
where $[\cdots]$ indicates the average over distinct samples. The typical DOS is
numerically calculated by the KPM. When the DOS is finite, the vanishing of the typical DOS
reflects the appearance of localized states. In Fig.~\ref{FigSM4}, we display both the DOS
and typical DOS in different phases. For $m_z=\pm \infty$, it is evident to see that
there exist regions around the band edges where the typical DOS vanishes while the DOS
is still finite, showing the localized characteristic in these regions; Yet, in other
regions, both the typical DOS and the DOS are finite, showing their extended feature.
This demonstrates the existence of the mobility edge where $\rho_{typ}/\rho$
vanishes. We also observe that the localized phase is more conspicuous for the positive
$E$ than the negative one, where the DOS is very small. This explains the clear presence
of the AAI for the negative $m_z$, but not for the positive one.
Now let us raise $m_z$ to $-4$, we observe that there still exists
a small region around zero and a region around other band edges which correspond
to a localized phase. As we increase $m_z$ further, the localized phase for the
former disappears while that for the latter persists. When $m_z\rightarrow \infty$,
both the DOS and the typical DOS are antisymmetric to the case when $m_z\rightarrow -\infty$.
As we decrease $m_z$ to $11.2$ and further to $10$, both the DOS and typical DOS
are very small. While the phase corresponds to a band insulator, the states around zero
energy are localized (which is also reflected by the LSR and IPR) despite the region being very
small.

\section{S-6. Effects of $R_c$ and stability against the on-site disorder}
\begin{figure}[t]
\includegraphics[width=6in]{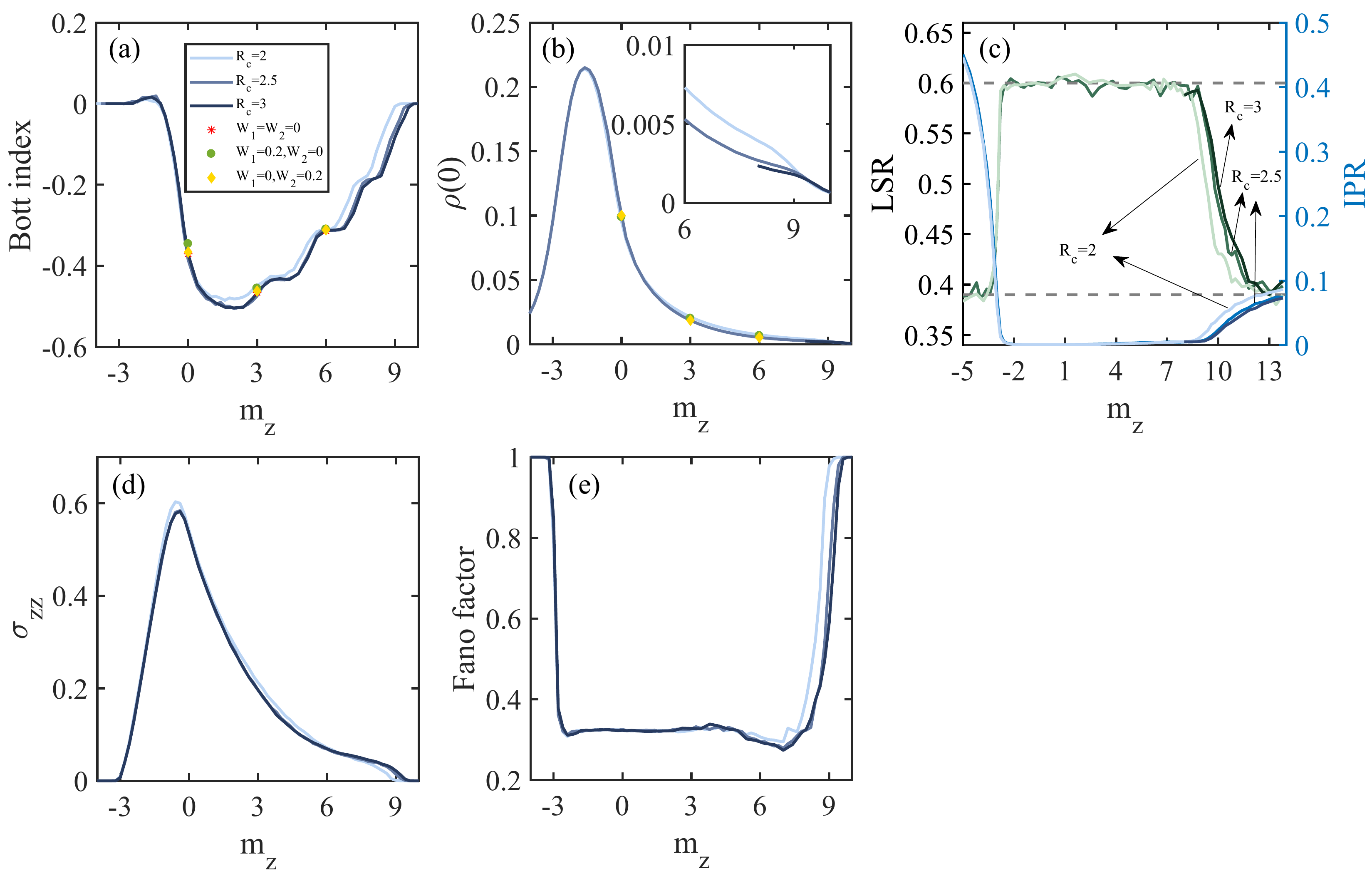}
\caption{(Color online) The Bott index for $L=16$ (a), the DOS at zero energy for $L=55$ and $N_c=2^{11}$ (b), the LSR and the IPR for
$L=24$ (c), the longitudinal conductivity (d) and the Fano factor (e) with respect to $m_z$ for three distinct $R_c$. In (a), the
red stars, green circles and yellow diamonds show the Hall conductivity for the disorder strength $W_1=W_2=0$,
$W_1=0.2,W_2=0$, and $W_1=0,W_2=0.2$
respectively. In (b), the green circles and yellow diamonds show the DOS at zero energy for $W_1=0.2,W_2=0$ and
 $W_1=0,W_2=0.2$, respectively, and the inset shows
the zoomed-in view of the DOS. In (d) and (e), the light and dark blue lines correspond to $L=24$ while
the other one $L=25$. Here, all samples are considered in a cubic box.}
\label{FigSM5}
\end{figure}

In this section, we discuss the effects of $R_c$ and the on-site disorder on our results. In Fig.~\ref{FigSM5}, we plot the Bott index, the DOS at zero energy, the LSR and the IPR, the longitudinal conductivity $\sigma_{zz}$ and the Fano factor for $R_c=2,2.5,3$. It clearly shows that $R_c$ has only quantitative effects on our results:
increasing $R_c$ from $2$ to $3$ only slightly shifts the phase boundary on the right side, while has vanishing effects
on the phase boundary on the left side. We note that the shift between $R_c=2.5$ and $3$ is very small.

To verify that our results are stable against the on-site disorder, we calculate the Hall conductivity and the DOS
at zero energy in the presence of the following term
\begin{equation}
H_D=\sum_{\bf x}\hat{c}_{\bf x}^\dagger[W_1V_1({\bf x})+W_2V_2({\bf x})\sigma_z]\hat{c}_{\bf x},
\end{equation}
where $V_1({\bf x})$ and $V_2({\bf x})$ are uniformly random variables chosen from $[-1,1]$. Fig.~\ref{FigSM5}(a-b) illustrates that
the presence of a weak disorder only has a very slight modification of the Hall conductivity and the DOS, suggesting
stability of our results against the on-site disorder.

\section{S-7. Experimental realization in electric circuits}
\begin{figure}[t]
\includegraphics[width=7in]{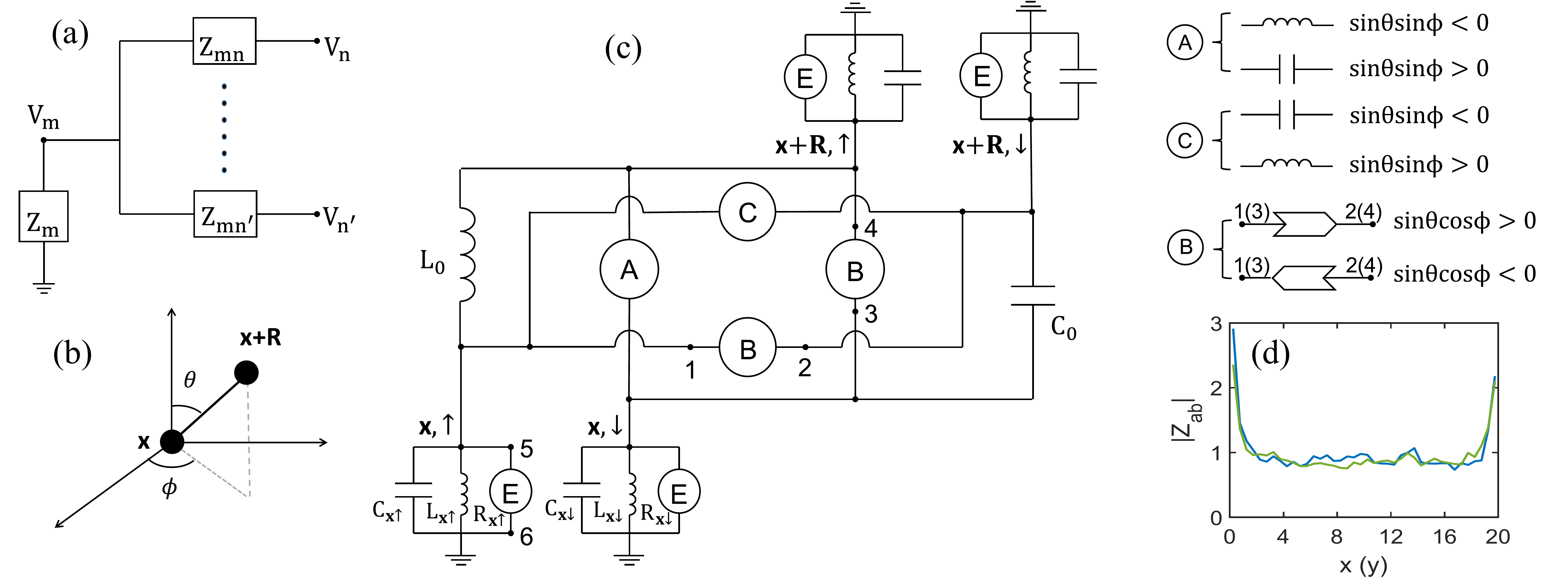}
\caption{(Color online) (a) Schematics of a simple electrical network. Electric circuits (c) between four nodes
(with spins) located at $\bf x$ and ${\bf x}+{\bf R}$ as shown in (b). $C_0$, $C_{{\bf x}\sigma}$, $L_0$,
and $L_{{\bf x}\sigma}$ denote the corresponding capacitances and inductances. The labeled circles represent electrical
elements which depend on the geometry between two nodes. For the circle labelled by B, it represents a negative impedance converter with current inversion (INIC)~\cite{Thomale2018arX,ChenBook}, the sign of the resistance depends on how
the INIC is connected. For instance, if $\sin\theta\cos\phi >0$, we require that the direction of the INIC is from point 1 to point 2, so that $I_{12}=-(V_1-V_2)/|R_B|$ corresponding to the negative resistance while
$I_{21}=(V_2-V_1)/|R_B|$ corresponding to the positive resistance. The electric element labelled by $E$ also depicts the
INIC with the corresponding resistance $R_{{\bf x}\sigma}$, the sign of which is dependent of the orientation of the INIC.
(d) The averaged two-node impedance versus the coordinate of each divided layer for $m_z=2$. For the blue (green) line, we divide the system into 40 layers perpendicular to the $x$ ($y$) direction and each pair of nodes are chosen randomly in each layer. The unit of the impedance is $\omega \bar{L}_0$.}
\label{FigSM6}
\end{figure}

In this section, we introduce a scheme (shown in Fig.~\ref{FigSM6}) to implement our Hamiltonian in electric circuits.
Let us consider an electrical network where the current flowing from node $m$ to node $n$
is denoted by $I_{mn}$ and the electric potential at each node $m$ is denoted by $V_m$. According to Kirchhoff's law,
\begin{equation}
I_m=\sum_n I_{mn}=\sum_{n}Y_{mn}(V_m-V_n)+Y_m V_m,
\end{equation}
where $Y_{mn}=1/Z_{mn}$ is the admittance between node $m$ and $n$ with $Z_{mn}$ being the corresponding impedance and
$Y_m=1/Z_m$ is the admittance between node $m$ and the ground as shown in Fig.~\ref{FigSM6}(a). We can write this equation
in a matrix form
\begin{equation}
I=JV,
\end{equation}
where $I=(\begin{array}{cccc}
           I_1 & I_2 & \cdots & I_M
         \end{array})^T
$ and
$V=(\begin{array}{cccc}
           V_1 & V_2 & \cdots & V_M
         \end{array})^T
$ with $M$ labelling the total number of nodes. Here, $J$ is the Laplacian acting as the Hamiltonian that can be used
to simulate our system. We note that such methods have been used to probe the SSH model~\cite{Thomale2018CP}, Weyl semimetals~\cite{Simon2019PRB} and higher topological insulators~\cite{Thomale2018NP}.

To implement our Hamiltonian, let us write the Hamiltonian as
\begin{equation}
H=\sum_{({\bf x},{\bf x}+{\bf R})} H({\bf x},{\bf x}+{\bf R})+\sum_{\bf x} H({\bf x}),
\end{equation}
where $H({\bf x},{\bf x}+{\bf R})$ depicts the hopping between two neighbor sites ${\bf x}$ and ${\bf x}+{\bf R}$
and $H({\bf x})$ the on-site term. We only need to construct the hopping between two sites and the on-site term and
all other connections can be built in a similar way. We propose an electric circuit shown in
Fig.~\ref{FigSM6}(c) which can be described by
\begin{equation}
J=i\omega [\mathcal{J}({\bf x},{\bf x}+{\bf R})+\mathcal{J}_0({\bf x},{\bf R})+\mathcal{J}({\bf x})],
\end{equation}
where
\begin{eqnarray}
\mathcal{J}({\bf x},{\bf x}+{\bf R})
=&&-C_0(R)\left[|\pi_{{\bf x},\uparrow}\rangle\langle \pi_{{\bf x}+{\bf R},\uparrow}|
-|\pi_{{\bf x},\downarrow}\rangle\langle \pi_{{\bf x}+{\bf R},\downarrow}|+
(i\sin\theta\cos\phi +\sin\theta\sin\phi)|\pi_{{\bf x}\uparrow}\rangle\langle \pi_{{\bf x}+{\bf R},\downarrow}|
\right. \nonumber \\
&&\left.+(i\sin\theta\cos\phi-\sin\theta\sin\phi)|\pi_{{\bf x},\downarrow}\rangle\langle \pi_{{\bf x}+{\bf R},\uparrow}|
+H.c.\right] \\
\mathcal{J}_0({\bf x},{\bf R})=&& -C_0(R)(1-\sin\theta\sin\phi-i\sin\theta\cos\phi)|\pi_{{\bf x},\uparrow}\rangle\langle \pi_{{\bf x},\uparrow}| \nonumber \\
&&-C_0(R)(-1+\sin\theta\sin\phi-i\sin\theta\cos\phi)|\pi_{{\bf x},\downarrow}\rangle\langle \pi_{{\bf x},\downarrow}| \\
\mathcal{J}({\bf x})=&& \sum_{\sigma=\uparrow,\downarrow}(C_{{\bf x},\sigma}-\frac{1}{\omega^2 L_{{\bf x},\sigma}}-i\frac{1}{\omega R_{{\bf x},\sigma}})|\pi_{{\bf x},\sigma}\rangle\langle \pi_{{\bf x},\sigma}|,
\end{eqnarray}
and $|\pi_{{\bf x},\uparrow}\rangle$ depicts a row vector with an entry corresponding to the node ${\bm \xi}=({\bf x},\sigma)$ being one and all other entries being zero,
and $C_0(R)=1/(\omega^2 L_0)=e^{\lambda(1-R)}/(2\omega^2 \bar{L}_0)$ with $\omega$ being the frequency of the alternating current. For the whole system, the contribution from $\mathcal{J}_0({\bf x},{\bf R})$ should be summed over
${\bf R}$. By appropriately tuning the circuit elements so that
$R_{{\bf x},\uparrow}=R_{{\bf x},\downarrow}=1/(\omega\sum_{\bf R}C_0(R)\sin\theta\cos\phi)$,
$C_{{\bf x}\uparrow}=1/(\omega^2 L_{{\bf x}\downarrow})$, $C_{{\bf x}\downarrow}=1/(\omega^2 L_{{\bf x}\uparrow})$
and $C_{{\bf x},\uparrow}-C_{{\bf x},\downarrow}=m_z/(\omega^2 \bar{L}_0)+\sum_{\bf R}C_0(R)(1-\sin\theta\sin\phi)$,
we achieve the expected Laplacian $J=iH/(\omega \bar{L}_0)$. To measure the surface states, we divide the system into
a number of layers and measure the impedance between two nodes in each layer. The averaged impedance
is given by
\begin{equation}
|Z_{ab}|= [|\sum_{n} \frac{|\Psi_{E_n,{\bm \xi}_a}-\Psi_{E_n,{\bm \xi}_b}|^2}{j_n}|],
\end{equation}
where $[\cdots]$ indicates the average over different pairs of two nodes and different samples
and $\Psi_{E_n,{\bm \xi}_a}$ is the ${\bm \xi}_a$ component of the eigenvector of $J$ corresponding
to the eigenvalue $j_n$.
Figure~\ref{FigSM6} plots the impedance in different layers, showing that the impedance
exhibits peaks around the boundaries, suggesting the presence of the surface states.

\end{widetext}
\end{document}